\documentclass[journal = cmatex, manuscript = article]{achemso}

\usepackage[version=4]{mhchem}
\usepackage{mathtools}
\usepackage{amsmath}
\usepackage{amssymb}
\usepackage[symbol]{footmisc}

\newcommand{\textsuper}[1]{$^{\text{#1}}$}
\newcommand{\textsub}[1]{$_{\text{#1}}$}

\title{Comparison of computational methods \\ for the electrochemical stability window \\ of solid-state electrolyte materials}

\author{Tobias Binninger}
\altaffiliation{Contributed equally to this work.}
\email{tobias.binninger.science@gmx.de}
\affiliation{IBM Research -- Zurich, S\"aumerstrasse 4, CH-8803 R\"uschlikon, Switzerland}
\author{Aris Marcolongo}
\altaffiliation{Contributed equally to this work.}
\affiliation{IBM Research -- Zurich, S\"aumerstrasse 4, CH-8803 R\"uschlikon, Switzerland}
\author{Matthieu Mottet}
\altaffiliation{Contributed equally to this work.}
\affiliation{IBM Research -- Zurich, S\"aumerstrasse 4, CH-8803 R\"uschlikon, Switzerland}
\alsoaffiliation{Theory and Simulation of Materials, \'Ecole Polytechnique F\'ed\'erale de Lausanne, CH-1015 Lausanne, Switzerland}
\author{Val\'ery Weber}
\affiliation{IBM Research -- Zurich, S\"aumerstrasse 4, CH-8803 R\"uschlikon, Switzerland}
\author{Teodoro Laino}
\affiliation{IBM Research -- Zurich, S\"aumerstrasse 4, CH-8803 R\"uschlikon, Switzerland}


\begin{document}

\maketitle

\begin{abstract}
Superior stability and safety are key promises attributed to all-solid-state batteries (ASSBs) containing solid-state electrolyte (SSE) compared to their conventional counterparts utilizing liquid electrolyte. To unleash the full potential of ASSBs, SSE materials that are stable when in contact with the low and high potential electrodes are required. The electrochemical stability window is conveniently used to assess the SSE--electrode interface stability. In the present work, we review the most important methods to compute the SSE stability window. Our analysis reveals that the stoichiometry stability method represents a bridge between HOMO--LUMO method and phase stability method (grand canonical phase diagram). Moreover, we provide computational implementations of these methods for SSE material screening. We compare their results for the relevant Li- and Na-SSE materials LGPS, LIPON, LLZO, LLTO, LATP, LISICON, and NASICON, and we discuss their relation to published experimental stability windows. 
\end{abstract}

\section{Introduction}

Solid-state electrolyte (SSE) materials, i.e. materials that provide ionic but no electronic conductivity, are key components of all-solid-state batteries (ASSBs), which are considered as promising candidates for next generation Li- and Na-ion batteries. State-of-the-art Li-ion batteries are based on liquid organic electrolytes, which are flammable and furthermore thermodynamically unstable against the low-potential electrode, e.g. metallic Li. Their operation is possible only because of the formation of a passivating layer between the electrode and the liquid electrolyte, termed the ``solid electrolyte interphase''\footnote{Strictly speaking, the orignial term was chosen to denote the solid ``phase'' of the passivating layer with a certain ionic conductivity, i.e. a ``solid electrolyte phase'' in between the electrode and the liquid electrolyte. Nowadays, however, also the term ``solid--electrolyte \emph{interface}'' is commonly used for SEI, i.e. the layer at the ``interface'' between the solid electrode and the liquid electrolyte. The use of the abbreviation SEI allows to neglect these subtle differences in terminology.} (SEI)~\cite{1979_Peled_J_Electrochem_Soc, 1983_Peled_J_Power_Sources}. In comparison to liquid electrolytes, many SSE materials offer superior stability against reduction and oxidation. Nevertheless, only few SSE materials provide true thermodynamic stability against the corresponding metallic anode~\cite{2015_Zhu_Mo_ACS_Appl_Mater_Interfaces, 2017_Lu_Ciucci_Chem_Mater}. Therefore, interfacial stability between the electrode and the solid-state electrolyte is a focus topic in current ASSB research~\cite{2016_Richards_Ceder_Chem_Mater, 2016_Li_Goodenough_PNAS}. For many SSE materials, a passivating interphase layer forms between the metallic low-potential electrode and the SSE~\cite{1974_Armstrong_J_Electroanal_Chem, 2015_Zhu_Mo_ACS_Appl_Mater_Interfaces, 2016_Richards_Ceder_Chem_Mater, 2016_Li_Goodenough_PNAS}, which consists of SSE decomposition products, 
similar to the SEI layer in liquid electrolyte Li-ion cells. Although such a ``metal--solid electrolyte interphase (MSEI) layer''~\cite{1974_Armstrong_J_Electroanal_Chem} can render an ASSB meta-stable due to passivation and also provide a certain ionic conductivity to sustain battery operation, the battery performance will be deteriorated because of the reduced ionic conductivity of the interphase layer. The same conclusion holds for passivating interphase layers between SSE and high-potential electrode~\cite{2017_Lu_Ciucci_Chem_Mater}. Thus, for ASSBs to unfold their full potential, stability of the SSE against both electrodes is desirable to avoid the formation of interphase layers.  

The interface stability properties of an SSE material in contact with an electrode are best described by its electrochemical stability window. We refer to ``electrochemical stability'' as the stability of an SSE material against \emph{reactions that contain transfer of atoms of the mobile species}, e.g. Li- or Na-atoms. It must be emphasized that only the transfer of \emph{neutral} atoms represents an SSE instability. In contrast, the transfer of mobile species ions, e.g. Li\textsuper{+} or Na\textsuper{+}, across the electrode--SSE interface is part of the normal battery operation and it does not alter the SSE composition or structure. 

\begin{figure}[t]
\includegraphics[width=0.4\columnwidth]{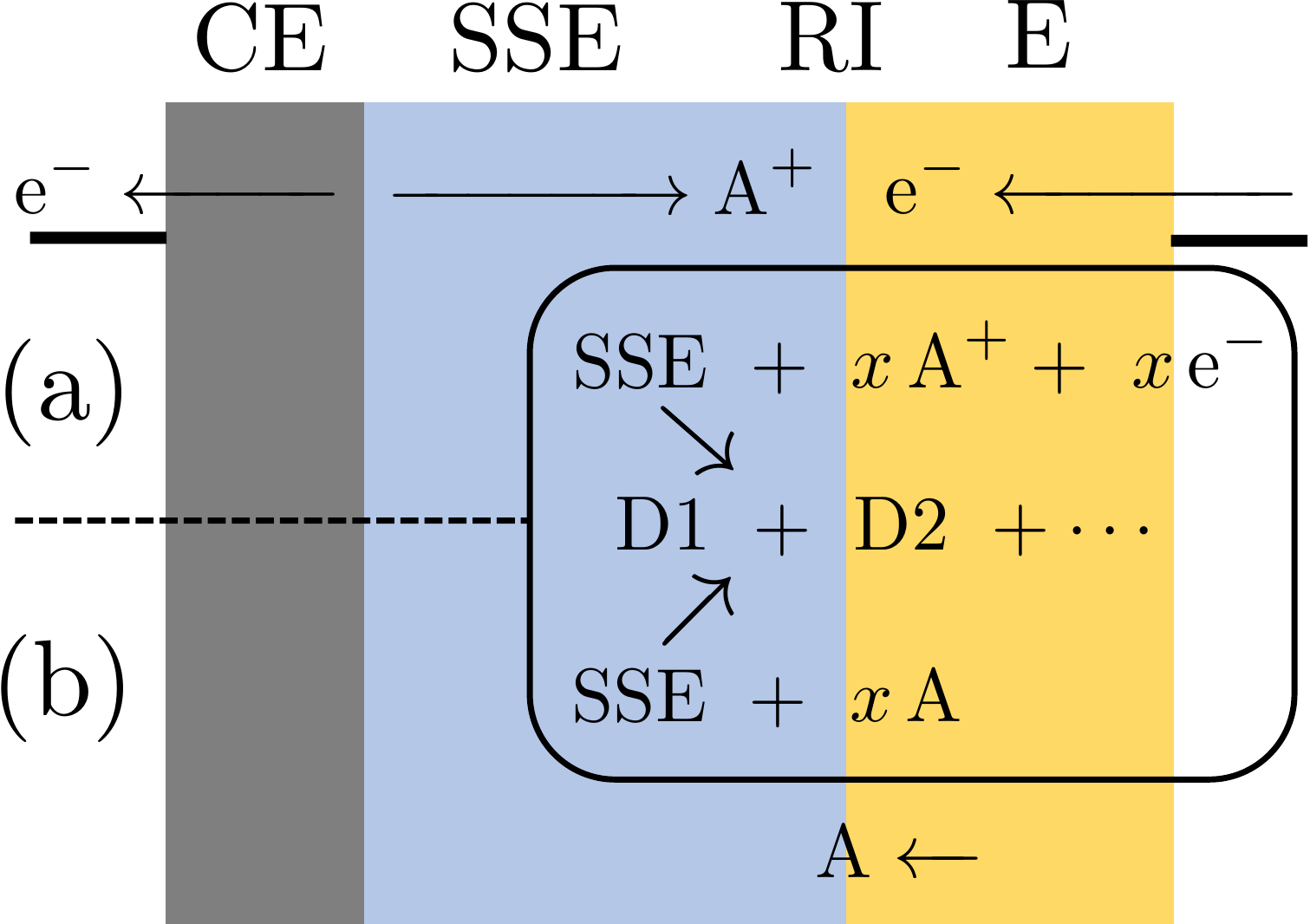}
\caption{The same SSE instability reaction in a closed-circuit cell (a) and open-circuit cell (b). CE = counter electrode, RI = reaction interface, E = electrode, \ce{A} = mobile species (\ce{Li}, \ce{Na}, \dots).}
\label{fig_SSE_instability_cell}
\end{figure}

This definition of ``electrochemical stability'' comprises both potential-driven SSE instability processes in a closed-circuit cell and direct reactions between SSE and electrodes under exchange of mobile species atoms that can also proceed in an open-circuit cell. In fact, both types of processes are equivalent: In an electrochemical process, the electrode represents an electron reservoir and electron transfer occurs across the electrode--electrolyte interface. If only electrons are transferred, the build-up of charge quickly stops the process. Thus, for an instability reaction to proceed, charge neutrality must be preserved i.e. the electronic charge transfer must be compensated by the movement of mobile ions to/from the instability reaction interface (RI), cf. Figure~\ref{fig_SSE_instability_cell}. In a closed-circuit electrochemical cell, Figure~\ref{fig_SSE_instability_cell}a, the compensating ions are supplied to interface RI through the SSE by the counter-reaction at the opposite electrode CE. However, in an ASSB the electro-active electrode materials are also ion reservoirs. Therefore, the compensating ions for the instability reaction can also be provided in an open-circuit ASSB from electrode E itself across the same electrode--SSE interface RI as the electrons. The combination of electron and ion transfer represents a transfer of \emph{neutral} mobile atoms between electrode and SSE, Figure~\ref{fig_SSE_instability_cell}b. 

According to this reasoning, a general electrochemical SSE instability reaction is written
\begin{align}
\label{eq_SSE_decomposition_overall}
\ce{SSE}\  +\ x\, \ce{A} \quad \longrightarrow \quad \ce{D}1\ +\ \ce{D}2\ +\ \cdots 
\end{align}
where \ce{A} = \ce{Li}, \ce{Na}, ... denotes the mobile species within an SSE material and \ce{D}1, \dots denotes different decomposition products. The coefficient $x$ can be positive or negative, depending on the direction of \ce{A}-transfer. Reaction~\eqref{eq_SSE_decomposition_overall} represents the overall cell reaction of Figure~\ref{fig_SSE_instability_cell}. The electrochemical character becomes more obvious from the corresponding half-cell reaction at the electrode--SSE interface RI,
\begin{align}
\label{eq_SSE_decomposition_half_cell}
\ce{SSE}\  +\ x\, \ce{A^+} +\ x\, \ce{e^-} \quad \longrightarrow \quad \ce{D}1\ +\ \ce{D}2\ +\ \cdots 
\end{align}
Thus, for $x>0$ the instability reaction is an SSE reduction, and for $x<0$ an SSE oxidation. If an \ce{A}-metal counter electrode is used, the counter half-cell reaction is simply given by 
\begin{align}
\label{eq_reference_reaction}
x\, \ce{A}\quad \longrightarrow\quad x\, \ce{A^+}\ +\ x\, \ce{e^-}
\end{align}
i.e. \ce{A}-metal oxidation for $x>0$ or plating for $x<0$. The equilibrium potential of the electrochemical SSE instability reaction~\eqref{eq_SSE_decomposition_half_cell} versus the reference electrode reaction~\eqref{eq_reference_reaction} is expressed by the Nernst equation in terms of the Gibbs free energy of the overall cell reaction~\eqref{eq_SSE_decomposition_overall}, cf. Supplementary Information for derivation,
\begin{align}
\label{eq_equilibrium_potential}
\Phi_{\text{eq}} = -\frac{\Delta G}{e\,x} = -\frac{1}{e}\,\left( \frac{G_{\ce{D}1} + G_{\ce{D}2} + \cdots - G_{\ce{SSE}}}{x} - G_{\ce{A}} \right) 
\end{align}
where $G_i$ denotes the Gibbs free energy of compound $i$. For $x>0$ (SSE reduction), reaction~\eqref{eq_SSE_decomposition_half_cell} proceeds if the SSE material is in contact with an electrode potential $\Phi < \Phi_{\text{eq}}$. Vice versa, for $x<0$ (SSE oxidation), reaction~\eqref{eq_SSE_decomposition_half_cell} proceeds if the SSE material is in contact with an electrode potential $\Phi > \Phi_{\text{eq}}$. The equilibrium potentials of all possible SSE instability reactions of type~\eqref{eq_SSE_decomposition_overall} yield an electrochemical series of SSE reduction and oxidation potentials. The limiting SSE reduction potential $\Phi_{\text{red}}$ is given by the maximum of all equilibrium potentials for reactions with $x>0$, $\Phi_{\text{red}} = \text{max}(\{\Phi_{\text{eq},i} \, | \, x_i>0 \})$, and the limiting SSE oxidation potential $\Phi_{\text{ox}}$ by the minimum of all equilibrium potentials for $x<0$, $\Phi_{\text{ox}} = \text{min}(\{\Phi_{\text{eq},j} \, | \, x_j<0 \})$. For electrode potentials $\Phi_{\text{red}} < \Phi < \Phi_{\text{ox}}$, no reduction or oxidation of the SSE occurs. Therefore, the potential range $[\Phi_{\text{red}}\, ,\, \Phi_{\text{ox}}]$ is the \emph{electrochemical stability window} of the SSE material.

Different computational methods have been applied to determine the electrochemical stability window. The positions of the electronic HOMO and LUMO states, i.e. the valence and conduction band edges, 
provide an estimate of the electrochemical stability window~\cite{2010_Goodenough_Kim_Chem_Mater, 2011_Ong_Ceder_Chem_Mater, 2012_Mo_Ceder_Chem_Mater}. In this approach, the electrode in contact with the SSE is considered as ``electron reservoir'' and only electron transfer between electrode and SSE is considered. The ``HOMO--LUMO method'' allows for a rather quick estimation of the width of the stability window in terms of the electronic band gap. However, its absolute position with respect to a reference electrode is difficult to determine with this methodology~\cite{2011_Ong_Ceder_Chem_Mater} because of the dipole at the electrode--SSE interface that shifts the relative positions of electronic states. Furthermore, because the electrode is assumed chemically inert, the HOMO--LUMO gap is considered only an upper bound for the electrochemical stability window~\cite{2012_Mo_Ceder_Chem_Mater, 2017_Lu_Ciucci_Chem_Mater}. 

The interface dipole problem is avoided by considering combined electron and ion transfer between SSE and electrode, i.e. transfer of neutral mobile Li- or Na-atoms, as described by reactions~\eqref{eq_SSE_decomposition_overall} and \eqref{eq_SSE_decomposition_half_cell}. The dipole does not affect the energies of charge-neutral states. Thus, the absolute position of the stability window w.r.t. a reference potential can be  conveniently determined from bulk computations. Two different cases are distinguished corresponding to two different methods to compute the stability window. 

The first method considers reactions of type~\eqref{eq_SSE_decomposition_overall} in the limit of small $x$ where the only product is the same SSE phase with a changed stoichiometry of the mobile species. In the following, we will refer to this method as ``stoichiometry stability method''. Such processes are similar to Li-insertion or Li-extraction reactions in electro-active electrode materials. Whereas in the latter case such reactions are part of the required function, for an SSE they represent an instability. This method was applied to compute the Li-insertion potentials of various garnet-type Li-SSE materials~\cite{2012_Nakayama_PCCP}, the Li-insertion and Li-extraction potentials of LGPS~\cite{2016_Bhattacharya_Wolverton_Carbon_Sci_Tech}, and the Na-extraction potentials of various Na-SSE materials~\cite{2017_Tian_Ceder_Energy_Environ_Sci}. A similar defect chemistry-derived perspective on SSE stability was adopted to investigate the instability of \ce{Li4P2S6} SSE against metallic Li~\cite{2018_Sadowski_Solid_State_Ionics}. Although this type of instability reaction is sometimes referred to as ``topochemical'' or ``topotactic'', we believe that this terminology must be used with care, because, strictly speaking, the latter terms denote an insertion reaction ``that results in significant structural modifications to the host'', as defined by IUPAC~\cite{1994_IUPAC}. In contrast, apart from some local relaxation, the SSE structure is usually considered unchanged when computing the insertion/extraction potentials of mobile species by filling/generating vacancies within the SSE host structure. 

On the contrary, the ``phase stability method'' considers reactions of type~\eqref{eq_SSE_decomposition_overall} that result in major decomposition of the SSE phase into several different product phases. This is typically done by constructing the grand canonical phase diagram of the SSE and identifying the critical values of the mobile species chemical potential that define the limiting SSE reduction and oxidation potentials~\cite{2008_Ong_Ceder_Chem_Mater, 2012_Mo_Ceder_Chem_Mater, 2015_Zhu_Mo_ACS_Appl_Mater_Interfaces}. Because the phase stability method takes into account a set of instability reactions that is disjoint from the stoichiometry stability method, the corresponding stability windows are complementary in the sense that the overall electrochemical stability window of an SSE material is the intersection of the stoichiometry stability window and the phase stability window. It is commonly assumed that the phase stability window is generally narrower than the stoichiometry stability window~\cite{2017_Tian_Ceder_Energy_Environ_Sci}, so that the overall stability window is determined by the former. However, to the best of our knowledge, no rigorous proof for this assumption exists to date. In contrast to the HOMO--LUMO method, both the stoichiometry stability method and the phase stability method yield a reliable absolute position of the electrochemical stability window w.r.t. a reference electrode potential, however at the expense of an increased computational effort.

In the present work, we analyse the relation between the different methods with a focus on the stoichiometry stability method. We provide computational implementations in an open-source repository~\cite{ZRL-AiiDA-toolbox} and we compare the results for a set of relevant Li- and Na-SSE materials.

\section{Methodology}

In the following, we indicate with \ce{A} = \ce{Li}, \ce{Na}, \dots, the mobile species within an SSE material. We first discuss the stoichiometry stability method, then its connection with the HOMO--LUMO method, and finally the phase stability method.

\subsection{Stoichiometry stability window}

We denote as \ce{A_{n}M} the stable composition for a unit cell of an SSE material, where the meaning of ``stable'' is further specified below. As explained in the Introduction, the stoichiometry stability method considers \ce{A}-insertion/extraction reactions to/from the SSE structure,
\begin{align}
\label{eq_vacancy_destruction}
\epsilon \ce{A} + \ce{A_{n}M} \quad \longrightarrow \quad \ce{A_{n+\epsilon}M}  \\[0.2cm]
\label{eq_vacancy_generation}
\ce{A_{n}M} \quad \longrightarrow \quad \ce{A_{n-\epsilon}M} + \epsilon \ce{A}
\end{align}
where $\epsilon$ is small compared to $n$. The critical difference between these instability processes and the transfer of \ce{A^+} ions during normal battery operation is the fact that the former change the SSE stoichiometry and the oxidation state of certain species in the \ce{M}-matrix. They represent redox-reactions between SSE and the respective electrode material: SSE reduction \eqref{eq_vacancy_destruction} and SSE oxidation \eqref{eq_vacancy_generation}.

For a general SSE stoichiometry \ce{A_{n+z}M} that deviates by $z$ from the stable composition, reactions \eqref{eq_vacancy_destruction} and \eqref{eq_vacancy_generation} read
\begin{align}
\label{eq_A_exchange_general}
\ce{A_{n+z}M}\ \pm\ \epsilon \ce{A}  \quad \longleftrightarrow \quad \ce{A}_{n+(z\pm\epsilon)}\ce{M}
\end{align}
The equilibrium potential $\Phi_{\text{eq}}$ of reaction~\eqref{eq_A_exchange_general} vs. the reference potential of the \ce{A}-metal oxidation process~\eqref{eq_reference_reaction} is a function of the deviation $z$ from the stable stoichiometry, and, according to equation~\eqref{eq_equilibrium_potential}, it is given by 
\begin{align}
\label{eq_equilibrium_potential_stoichio}
\Phi_{\text{eq}}(z) & = -\frac{1}{e}\,\left( \frac{G_{\ce{A}_{n+(z\pm\epsilon)}\ce{M} } - G_{\ce{A_{n+z}M}}}{\pm\epsilon} - G_{\ce{A}} \right) \\[0.2cm]
\label{eq_equilibrium_potential_stoichio_chem_pot}
& = -\frac{1}{e}\,\left( \mu^{\ce{A}}_{\text{SSE}}(z) - \mu^{\ce{A}}_{\ce{A}} \right)
\end{align}
Here, $\mu^{\ce{A}}_{\ce{A}} = G_{\ce{A}}$ is the chemical potential of \ce{A}-metal, i.e. its Gibbs free energy per atom, and $\mu^{\ce{A}}_{\text{SSE}}(z)$ corresponds to the chemical potential of neutral species \ce{A} in the SSE composition \ce{A_{n+z}M},
\begin{align}
\label{eq_chemical_potential}
\mu^{\ce{A}}_{\text{SSE}}(z) = \lim_{\epsilon \rightarrow 0} \frac{G_{\ce{A}_{n+(z+\epsilon)}\ce{M} } - G_{\ce{A_{n+z}M}}}{\epsilon} = \frac{\text{d}G}{\text{d}z}
\end{align}
which is equal to the derivative of the Gibbs free energy $G(z) \coloneqq G_{\ce{A_{n+z}M}}$ as a function of the stoichiometry deviation $z$. Equation~\eqref{eq_equilibrium_potential_stoichio} is identical to the general relation for the \ce{A}-insertion potential of electro-active battery materials~\cite{2010_Chevrier_Ceder_PRB, 2014_Saubanere_Doublet_Nat_Comm, 2016_Urban_Ceder_NPJ_Compu_Mat}.

In general, the stable SSE composition \ce{A_{n}M} corresponds to a stoichiometry where all constituting ions are formally in an electronic closed-shell configuration. Consequently, SSE materials have a band gap $E^{\text{LUMO}}_{\ce{A_{n}M}} - E^{\text{HOMO}}_{\ce{A_{n}M}} > 0$ and are electronic isolators as required for their function as electrolyte. There exists a qualitative difference between adding \ce{A} to and extracting \ce{A} from the stable composition \ce{A_{n}M}: The electrons of added \ce{A} populate LUMO states of \ce{A_{n}M}, whereas the electrons of extracted \ce{A} are removed from HOMO states of \ce{A_{n}M}. Therefore, the stable stoichiometry $n$ separates two distinct energetic manifolds with the consequence that the chemical potential $\mu^{\ce{A}}_{\text{SSE}}(z)$ has a discontinuity at $z=0$ with $\lim_{z\rightarrow 0^+} \mu^{\ce{A}}_{\text{SSE}}$ being strictly larger than $\lim_{z\rightarrow 0^-} \mu^{\ce{A}}_{\text{SSE}}$. Consequently, we obtain two different equilibrium potentials: The potential of A-insertion (SSE reduction) into the stable SSE stoichiometry \ce{A_{n}M}, 
\begin{align}
\label{eq_phi_red_stoi}
\Phi_{\text{red}}^{\text{stoi}} = -\frac{1}{e}\,\left( \lim_{z\rightarrow 0^+} \mu^{\ce{A}}_{\text{SSE}} - \mu^{\ce{A}}_{\ce{A}} \right) = -\frac{1}{e}\,\left( \left.\frac{\text{d}G}{\text{d}z^+}\right|_{z=0} - \mu^{\ce{A}}_{\ce{A}} \right)
\end{align}
and the potential of A-extraction from the stable SSE stoichiometry \ce{A_{n}M},
\begin{align}
\label{eq_phi_ox_stoi}
\Phi_{\text{ox}}^{\text{stoi}} = -\frac{1}{e}\,\left( \lim_{z\rightarrow 0^-} \mu^{\ce{A}}_{\text{SSE}} - \mu^{\ce{A}}_{\ce{A}} \right) = -\frac{1}{e}\,\left( \left.\frac{\text{d}G}{\text{d}z^-}\right|_{z=0} - \mu^{\ce{A}}_{\ce{A}} \right)
\end{align}
For electrode potentials $\Phi_{\text{red}}^{\text{stoi}} < \Phi < \Phi_{\text{ox}}^{\text{stoi}}$, the SSE stoichiometry is stable and fixed at \ce{A_{n}M}, and the valencies of the SSE constituent ions remain constant. Therefore, we call the potential range $[\Phi_{\text{red}}^{\text{stoi}}\, ,\, \Phi_{\text{ox}}^{\text{stoi}}]$ the \emph{stoichiometry stability window} of the SSE material. 

Two simple models that grasp the essential behaviour of the SSE \ce{A}-stoichiometry as a function of the electrode potential are presented in the Supplementary Information. Similar models exist for the Li-insertion into electro-active materials~\cite{1981_Raistrick_Solid_State_Ionics}. The resulting SSE stoichiometry vs. electrode potential curves are plotted in Figure~\ref{fig_stoichiometry_window_model} for generic values of the model parameters given in the Supplementary Information. Within $[\Phi_{\text{red}}^{\text{stoi}}\, ,\, \Phi_{\text{ox}}^{\text{stoi}}]$, the \ce{A}-stoichiometry is fixed at $n$. Outside the stoichiometry stability window, the SSE stoichiometry quickly changes by inserting (for $\Phi < \Phi_{\text{red}}^{\text{stoi}}$) or extracting (for $\Phi > \Phi_{\text{ox}}^{\text{stoi}}$) \ce{A}-atoms. In this region, the behaviour of the \ce{A}-stoichiometry is determined by the configurational entropy of \ce{A}-site occupation, which produces a steep step of the stoichiometry within a narrow potential range: The SSE turns into an electro-active material with a plateau of the equilibrium potential as a function of \ce{A}-stoichiometry, cf. Figure~\ref{fig_stoichiometry_window_model} rotated by $90^{\circ}$.  

\begin{figure}[t]
\includegraphics[width=0.6\columnwidth]{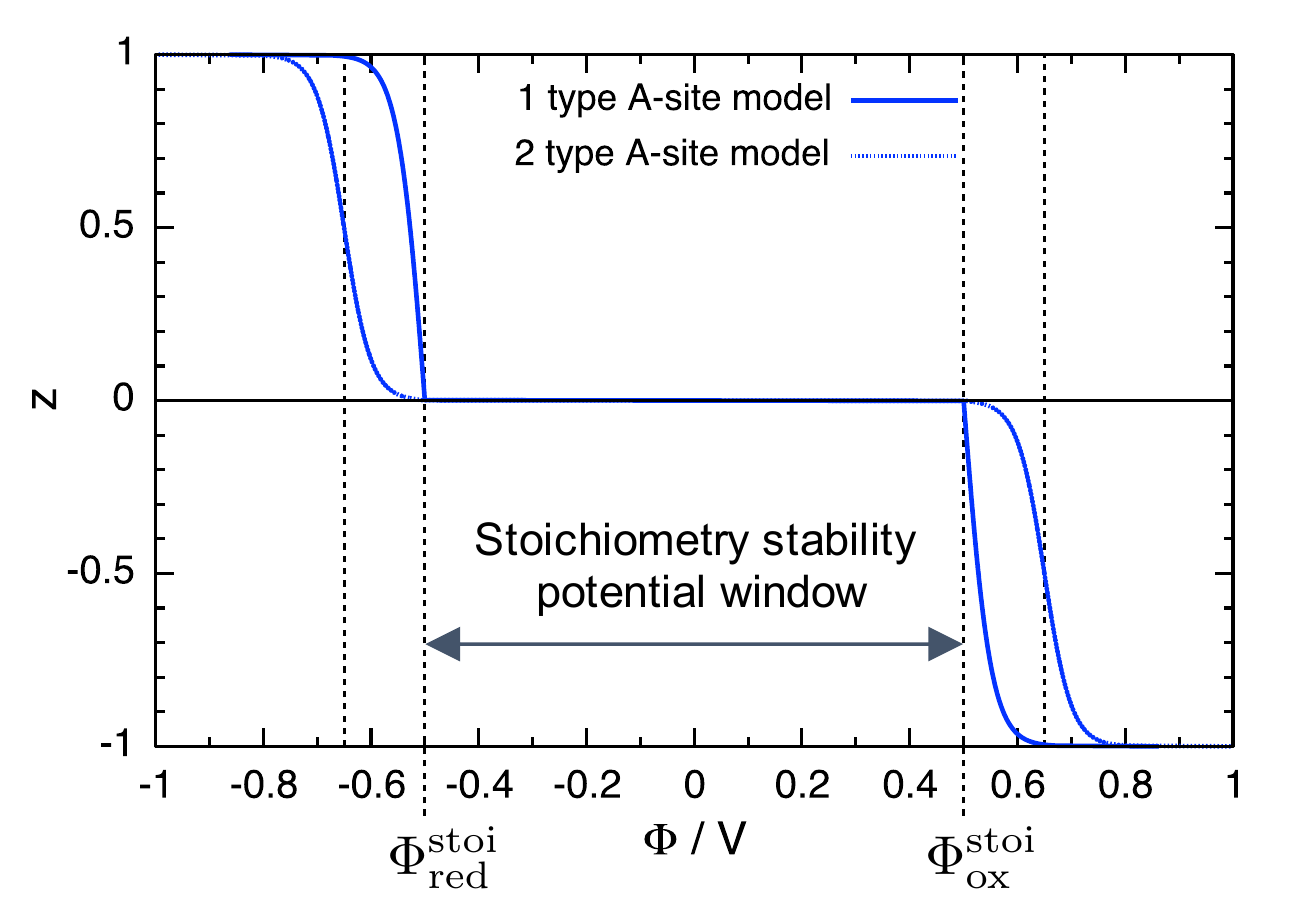}
\caption{The stoichiometry deviation $z$ in the general SSE composition \ce{A_{n+z}M} as a function of the electrode potential $\Phi$ as derived from two simple models for generic values of the model parameters, cf. Supplementary Information.}
\label{fig_stoichiometry_window_model}
\end{figure}

\subsubsection{Stoichiometry stability window: Relevance for ASSB application}

The stable stoichiometry \ce{A_{n}M} corresponds to an electronic insulator with a band gap. However, addition of \ce{A} to the material introduces electrons into the conduction band. Similarly, extraction of \ce{A} generates electron holes in the valence band. Therefore, even small stoichiometry changes of the SSE material can significantly increase its electronic conductivity with the consequence of an electronic short circuit between anode and cathode via the SSE. Electronic short-circuit of the electrodes causes self-discharge of an ASSB via combined electron and ion migration through the SSE. Of course, the extent of electronic conductivity increase due to \ce{A}-stoichiometry changes is strongly material-specific, because it depends on whether the additional electronic charge carriers occupy mobile states within the conduction or valence band, or whether they get trapped by in-gap states. 

Furthermore, local changes of the SSE stoichiometry at one or both of the SSE--electrode interfaces can result in a chemical short circuit between the electrodes and a diffusion-driven self-discharge of the ASSB. At both contact interfaces, the SSE stoichiometry equilibrates with the respective electrode potential. If both the anode potential $\Phi_A$ and the cathode potential $\Phi_C$ lie inside the potential window $[\Phi_{\text{red}}^{\text{stoi}}\, ,\, \Phi_{\text{ox}}^{\text{stoi}}]$, the SSE stoichiometry deviation $z$ is equal to zero everywhere, cf. Figure~\ref{fig_stoichiometry_window_model}. If, however, $\Phi_A < \Phi_{\text{red}}^{\text{stoi}}$ and/or $\Phi_C > \Phi_{\text{ox}}^{\text{stoi}}$, the SSE stoichiometry deviation $z_A$ at the anode contact interface will be different from the value $z_C$ at the cathode contact interface, with $z_A > z_C$. This causes an \ce{A}-concentration gradient across the SSE layer and triggers diffusion of neutral \ce{A}-species from the anode side to the cathode side. This mechanism can also proceed at open-circuit and result in self-discharge of the ASSB via mobile species diffusion.

The speed of the diffusion self-discharge depends critically on the value of the diffusion coefficient of \emph{neutral} \ce{A} within the non-stoichiometric \ce{A_{n+z}M} SSE phase. We consider 
an ASSB with an SSE layer thickness of $d_{\text{SSE}}=100\,\mu$m. Assuming a small relative SSE stoichiometry gradient of $\Delta z/n = 0.05$ between the anode interface and the cathode interface, we estimate a gradient of mobile ion density $\Delta n_A/d_{\text{SSE}} = n_A \, (\Delta z/n)/d_{\text{SSE}} =  5\times 10^{21}\,$cm\textsuper{-4} for a typical mobile ion density $n_A = 10^{21}\,$cm\textsuper{-3}. According to Fick's law of diffusion, an \ce{A}-diffusion coefficient $D_A = 10^{-7}\,$cm\textsuper{2}s\textsuper{-1} yields an \ce{A}-diffusion flux of $|j_A| = D_A \Delta n_A / d_{\text{SSE}} = 5\times 10^{14}\,$cm\textsuper{-2}s\textsuper{-1}, which is equivalent to an electrical self-discharge current of $|j_{el}| = e\,|j_A| \approx 0.08\,$mA\,cm\textsuper{-2}. A typical Li-ion battery contains approx. $20\,$mg\,cm\textsuper{-2} active cathode material with specific capacitance of approx. $150\,$mAh/g, yielding an area-specific battery capacitance of approx. $3\,$mAh\,cm\textsuper{-2}. Such a battery would be entirely self-discharged within approx. $1.5\,$days. Even for $D_A = 10^{-8}\,$cm\textsuper{2}s\textsuper{-1}, a complete self-discharge time of approx. two weeks is obtained. 

Values of $10^{-8}$--$10^{-7}\,$cm\textsuper{2}s\textsuper{-1} are common for the \emph{ionic} diffusion coefficient in SSE materials with high ionic conductivity. However, ionic diffusion of \ce{A^+} is different from the diffusion of \emph{neutral} \ce{A} considered here: The latter is equivalent to a parallel diffusion of both \ce{A^+} ions and electrons as described by Maier~\cite{1993_Maier_J_Am_Ceramic_Soc_I}. If the electronic mobility is large, the diffusion coefficient of neutral \ce{A} will be essentially equal to the ionic one. This is clearly not fulfilled for an SSE material in its stable stoichiometry which is an electronic isolator. But, as discussed above, the electronic conductivity of the non-stoichiometric SSE can be significantly increased. Thus, whether diffusion self-discharge can be relevant depends on the SSE material at hand and on the electronic conductivity properties of its non-stoichiometric phase. 

\subsubsection{Stoichiometry stability window: Computational Implementation}

We implemented the stoichiometry stability method in the Python workflow environment AiiDA~\cite{AiiDA} utilizing methods of the Python Materials Genomics (pymatgen) module~\cite{pymatgen}. We provide the stoichiometry stability plug-in in the ZRL-AiiDA-toolbox repository on GitHub~\cite{ZRL-AiiDA-toolbox}. 

The potential limits $\Phi_{\text{red}}^{\text{stoi}}$ and $\Phi_{\text{ox}}^{\text{stoi}}$ of the stoichiometry stability window are given by equations~\eqref{eq_phi_red_stoi} and \eqref{eq_phi_ox_stoi}, respectively. We choose a large supercell of the SSE structure with a stable stoichiometry \ce{A_{N}M} and we approximate the derivatives of the Gibbs free energy by the energy differences of adding/removing one \ce{A} atom to/from the supercell, respectively, $\left.\frac{\text{d}G}{\text{d}z^{\pm}}\right|_{z=0} \approx \pm(E_{\ce{A_{N\pm 1}M}}^{\text{min}} - E_{\ce{A_{N}M}}^{\text{min}})$. Here, we neglect the $pV$ term, entropic contributions, and thermal contributions to the internal energy, which is well justified if an error of $\pm 0.1\,$V is acceptable on the calculated stability potential window. SSE materials contain both occupied and unoccupied \ce{A}-sites. Every distribution $i$ of the \ce{A}-atoms over the available \ce{A}-sites has a distinct energy. We take the minimum over all configurations, i.e. the configurational ground-state energies $E_{\ce{A_{N}M}}^{\text{min}} = \text{min}(\{ E_{\ce{A_{N}M}}^{i}\ |\ i \in \text{conf}_{\ce{A_{N}M}}\})$ and analogously $E_{\ce{A_{N\pm 1}M}}^{\text{min}}$. Again, the contribution of thermally activated configurations can be neglected if an error of $\pm 0.1\,$V is acceptable. Then the potential limits read
\begin{align}
\label{eq_lower_potential_limit_method}
\Phi_{\text{red}}^{\text{stoi}} = -\frac{1}{e}(E_{\ce{A_{N+1}M}}^{\text{min}} - E_{\ce{A_{N}M}}^{\text{min}} - E^{\ce{A}}_{\ce{A}})  \\[0.2cm]
\label{eq_upper_potential_limit_method}
\Phi_{\text{ox}}^{\text{stoi}} = -\frac{1}{e}(E_{\ce{A_{N}M}}^{\text{min}} - E_{\ce{A_{N-1}M}}^{\text{min}} - E^{\ce{A}}_{\ce{A}})
\end{align}
Here, the chemical potential of \ce{A}-metal $\mu^{\ce{A}}_{\ce{A}}$, which is equal to the Gibbs free energy per atom, is approximated by the energy per atom of the relaxed \ce{A}-metal structure, $\mu^{\ce{A}}_{\ce{A}} \approx E^{\ce{A}}_{\ce{A}}$. 

\paragraph{Monte Carlo sampling.} An efficient sampling method is required to find the configurational ground-state energies, because a complete sampling of all configurations is not possible, and because the energies can vary significantly between different configurations. We use a Monte Carlo sampling algorithm based on purely electrostatic Ewald energies to identify a number of candidate configurations for the configurational ground-state. A random initial distribution of \ce{A}-atoms over the available \ce{A}-sites is generated according to the occupancies defined in the SSE material crystallographic file. Every Monte Carlo step consists of swapping a randomly chosen \ce{A}-atom to an unoccupied \ce{A}-site. The swap is accepted according to the Metropolis criterion with a probability $p = \text{min}(1, \exp(-\Delta E/(k_B T)))$, where $\Delta E$ is the Ewald energy difference of the swap, $k_B$ is the Boltzmann constant, and $T$ is the temperature chosen. The corresponding Ewald energies are computed from the formal charges of each species given in the crystallographic file using the pymatgen Python module~\cite{pymatgen}. The Monte Carlo process runs until equilibration is reached. 

We then use the final configurations of approximately 30 independent Monte Carlo runs to compute their density functional theory (DFT) energies including structural relaxation. Finally, the minimum of the relaxed DFT energies is determined. The entire process is performed separately for the three compositions \ce{A_{N}M}, \ce{A_{N+1}M}, and \ce{A_{N-1}M} to determine $E_{\ce{A_{N}M}}^{\text{min}}$, $E_{\ce{A_{N+1}M}}^{\text{min}}$, and $E_{\ce{A_{N-1}M}}^{\text{min}}$, respectively.

The sampling method is based on the assumption that configurations with minimum DFT energies also have electrostatic Ewald energies close to the minimum, which has also been used by other authors~\cite{2012_Mo_Ceder_Chem_Mater}. To further verify this assumption, we plot Ewald energies vs. DFT energies in Supplementary Information Figure~S2 for the SSE materials LGPS, LIPON, LLZO, LLTO, LATP, LISICON, and NASICON. For each of the compositions \ce{A_{N-1}M}, \ce{A_{N}M}, and \ce{A_{N+1}M}, we selected 100 random configurations. A linear correlation is observed in most cases. Typical slopes of the order of 10 indicate the dielectric screening properties of the materials, which are neglected in Ewald energies. Based on the slopes and the scattering of the energy--energy plots, we chose a temperature of 5000\,K for the Monte Carlo runs.

\paragraph{DFT details.} DFT computations were performed using the Quantum ESPRESSO software package~\cite{2009_Quantum_ESPRESSO_J_Phys_Cond_Matt}. The generalized gradient approximation (GGA) of the exchange-correlation functional in PBE form~\cite{1996_Perdew_Burke_Ernzerhof_PRL} was used along with pseudopotentials from the SSSP Efficiency library\cite{2018_Prandini_SSSP_NPJ_Comp_Mater, 2016_Lejaeghere_Science}. We chose the double of the cutoff values suggested by SSSP for wavefunctions and density. Because of the usage of supercells, $k$-point sampling was restricted to the $\Gamma$-point. The convergence of the results was confirmed by certain computations using a 2x2x2 $k$-point mesh. If possible, the magnetization, i.e. the population difference $\Delta n$ between up- and down-spin, was allowed to relax. In case of failure, it was fixed at $\Delta n=0$ for \ce{A_{N}M} (closed-shell) and at $\Delta n=1$ for \ce{A_{N$\pm$ 1}M} (single unpaired electron). We further used Marzari-Vanderbilt smearing~\cite{1999_Marzari_PRL} with 0.005\,Ry.

We also tested the influence of hybrid functionals using HSE06~\cite{2006_HSE06_J_Chem_Phys} with an exact exchange fraction of 0.25 together with SG15 ONCV pseudopotentials~\cite{2013_Hamann_PRB_ONCV_pseudo, 2015_Schlipf_Gygi_Comp_Phys_Comm_ONCV_pseudo}. We used a wavefunction cutoff of 100\,Ry for LATP and LLTO,  and 60\,Ry for LLZO (because of its very large cell). The density cutoff was four times the wavefunction cutoff.

\subsubsection{Stoichiometry stability window and HOMO--LUMO method}

The SSE stoichiometry stability window is determined by the discontinuity in the chemical potential of neutral \ce{A} at the stable stoichiometry. This discontinuity originates from the electronic HOMO--LUMO gap (or band gap). Therefore, a very close relation exists between the stoichiometry stability method and the HOMO--LUMO method. In fact, the chemical potential of neutral \ce{A} in the SSE phase, cf. equation~\eqref{eq_chemical_potential}, is equal to the sum of \ce{e^-} and \ce{A^+} chemical potentials,
\begin{align}
\label{eq_chemical_potential_sum}
\mu^{\ce{A}}_{\text{SSE}}\ =\ \frac{\text{d}G}{\text{d}z}\ =\ \frac{\partial G}{\partial z_{\ce{e^-}}} + \frac{\partial G}{\partial z_{\ce{A^+}}}\ =\ \mu^{\ce{e^-}}_{\text{SSE}} + \mu^{\ce{A^+}}_{\text{SSE}}
\end{align}
where $z$, $z_{\ce{e^-}}$, and $z_{\ce{A^+}}$ are the changes in neutral \ce{A}, \ce{e^-}, and \ce{A^+} numbers, respectively. The ionic chemical potential $\mu^{\ce{A^+}}_{\text{SSE}}$ essentially contains the interaction between \ce{A^+} and the SSE crystal field, i.e. the Madelung potential of the \ce{A}-sites~\cite{2014_Saubanere_Doublet_Nat_Comm}. Therefore, we assume that the ionic chemical potential is continuous at the stable SSE composition, $\lim_{z\rightarrow 0^-} \mu^{\ce{A^+}}_{\text{SSE}} = \lim_{z\rightarrow 0^+} \mu^{\ce{A^+}}_{\text{SSE}}$. From equations~\eqref{eq_phi_red_stoi} and ~\eqref{eq_phi_ox_stoi}, we then obtain for the \emph{width of the stoichiometry stability window}
\begin{align}
\nonumber
\Delta\Phi^{\text{stoi}} = \Phi_{\text{ox}}^{\text{stoi}} - \Phi_{\text{red}}^{\text{stoi}}\ & =\ -\frac{1}{e}\,\left( \lim_{z\rightarrow 0^-} \mu^{\ce{A}}_{\text{SSE}} - \lim_{z\rightarrow 0^+} \mu^{\ce{A}}_{\text{SSE}} \right) \\[0.2cm]
\nonumber
& =\ -\frac{1}{e}\,\left( \lim_{z\rightarrow 0^-} \mu^{\ce{e^-}}_{\text{SSE}} - \lim_{z\rightarrow 0^+} \mu^{\ce{e^-}}_{\text{SSE}} \right) \\[0.2cm]
\label{eq_width_stoi_window_IP_EA}
& =\ \frac{1}{e}\,\left( \text{IP}_{\text{SSE}} - \text{EA}_{\text{SSE}}  \right)
\end{align}
where we used the definitions of ionization potential $\text{IP} = -\lim_{z\rightarrow 0^-} \mu^{\ce{e^-}}_{\text{SSE}}$ and electron affinity $\text{EA} = -\lim_{z\rightarrow 0^+} \mu^{\ce{e^-}}_{\text{SSE}}$~\cite{1982_Perdew_PRL}. We see that the width of the stoichiometry stability window is equal to the fundamental gap of the SSE~\cite{1983_Perdew_PRL}. In practice, the potential limits are computed from finite \ce{A} number differences as in equations~\eqref{eq_lower_potential_limit_method} and \eqref{eq_upper_potential_limit_method}. Then, relation~\eqref{eq_width_stoi_window_IP_EA} remains approximately valid up to energetic differences between addition/removal of one \ce{A^+} ion and interaction terms between the added/removed \ce{A^+} and \ce{e^-}. 

An estimation of the fundamental gap from the HOMO--LUMO gap of Kohn-Sham DFT orbitals is non-trivial. The HOMO--LUMO gap $\Delta E_{\text{HL}}^{\text{N}} = E_{\text{LUMO}}^{\text{N}} - E_{\text{HOMO}}^{\text{N}}$ of the system with stable \ce{A}-stoichiometry N gives a very poor estimate of the fundamental gap, because of a discontinuity in the DFT exchange-correlation (xc) potential~\cite{1983_Perdew_PRL, 1985_Sham_PRB, 2010_Tsuneda_J_Chem_Phys}. Long-range corrected DFT xc-functionals are known to approximately fulfil Koopmans' theorem and yield much better estimates of IP and EA from Kohn-Sham HOMO and LUMO energies~\cite{2010_Tsuneda_J_Chem_Phys}. We therefore also compute $\Delta E_{\text{HL}}^{\text{N}}$ for a few materials using hybrid functional DFT, cf. details given above for the stoichiometry stability method.

\subsection{Phase stability window}

The electrochemical stability of an SSE material against major decomposition of the SSE phase is typically assessed by computing its \ce{A}-grand canonical phase diagram~\cite{2008_Ong_Ceder_Chem_Mater, 2012_Mo_Ceder_Chem_Mater, 2015_Zhu_Mo_ACS_Appl_Mater_Interfaces}. The electro-active material of the electrode represents an \ce{A}-reservoir. The corresponding \ce{A}-chemical potential $\mu^{\ce{A}}$ is fixed by the electrode potential $\Phi$ via the relation $\Phi = -(\mu^{\ce{A}} - \mu^{\ce{A}}_{\ce{A}})/e$, which is analogous to equation~\eqref{eq_equilibrium_potential_stoichio_chem_pot}, with the \ce{A}-metal reference chemical potential $\mu^{\ce{A}}_{\ce{A}}$.

We compute the phase stability window of an SSE in an equivalent, but slightly different way. For an SSE material with composition \ce{A_{a}B_{b}C_{c}D_{d}}, we consider all known compounds that contain one or several of the elements \ce{A}, \ce{B}, \ce{C}, or \ce{D}, and we construct all possible SSE decomposition reactions of type~\eqref{eq_SSE_decomposition_overall}. For each decomposition reaction $i$, we then compute the corresponding equilibrium potential $\Phi_{\text{eq},i}$ from equation~\eqref{eq_equilibrium_potential}. Tolerating errors of the order of 0.1\,V on the equilibrium potentials, we approximate the Gibbs free energies $G$ by the energies $E$ after structural relaxation. The equilibrium potentials are grouped according to the type of reaction, i.e. either SSE reduction if $x_i>0$, or SSE oxidation if $x_i<0$. If $x_i=0$, the decomposition reaction is not electrochemical, because it does not involve electron and ion transfer, cf. Introduction. Then, the potential limits of the phase stability window $[\Phi_{\text{red}}^{\text{phase}}\, ,\, \Phi_{\text{ox}}^{\text{phase}}]$ are given by $\Phi_{\text{red}}^{\text{phase}} = \text{max}(\{\Phi_{\text{eq},i} \, | \, x_i>0 \})$ and $\Phi_{\text{ox}}^{\text{phase}} = \text{min}(\{\Phi_{\text{eq},j} \, | \, x_j<0 \})$. 

The combinatorics of relevant decomposition reactions $\ce{SSE} \rightarrow d_1\,\ce{D}1 + \cdots + d_m\,\ce{D}m - x\, \ce{A}$ are restricted, because the maximum number $m$ of decomposition products in a given reaction is equal to the number of distinct elements in the SSE minus one. We represent the SSE and all products by composition vectors with one dimension for each distinct element, e.g. an SSE material \ce{A_{a}B_{b}C_{c}D_{d}} by the vector $(a, b, c, d)$. Then the reaction stoichiometry coefficients $d_1$, \dots, $d_m$, and $-x$ are equal to the expansion coefficients of the SSE vector $(a, b, c, d)$ in a basis given by the set of product vectors, including the vector $(1, 0, 0, 0)$ representing \ce{A}-metal. The total number of basis vectors is equal to the dimension, i.e. the number of distinct elements. Because the vector representing \ce{A}-metal is always present, the number $m$ of additional basis vectors is equal to the number of distinct elements minus one. Any reaction with more than $m$ decomposition products can be written as a superposition of such ``basic'' reactions, which proceed independently. Therefore, it is sufficient to consider only basic SSE decomposition reactions. This restriction is analogous to Gibbs' phase rule that restricts the number of coexisting equilibrium phases. Furthermore, only decomposition reactions with stoichiometry coefficients $d_j \geq 0$ must be taken into account, because the only reactants are the SSE and, in case of SSE reduction, the \ce{A}-metal reference.

\paragraph{Computational details.} We provide our implementation of the phase stability method as an AiiDA~\cite{AiiDA} plug-in in the ZRL-AiiDA-toolbox repository on GitHub~\cite{ZRL-AiiDA-toolbox}. For every investigated SSE material, we considered all possible decomposition products included in the Inorganic Crystal Structure Database (ICSD)~\cite{1983_ICSD, 2002_ICSD}. For every product, we created a supercell and computed the DFT energy after structural relaxation. The same DFT parameters were used as given above for the stoichiometry stability method. For products with partial occupancies, we used the Ewald energy-based Monte Carlo algorithm to generate a favourable configuration, cf. implementation of the stoichiometry stability method. 

\section{Results and Discussion}

\paragraph{Comparison of stoichiometry stability and phase stability windows.}

We compare our computed stoichiometry stability windows with published~\cite{2015_Zhu_Mo_ACS_Appl_Mater_Interfaces} phase stability windows for a set of important Li-SSE materials: \ce{Li10GeP2S12} (LGPS), \ce{O}-doped \ce{Li2PO2N} (LIPON), \ce{Li7La3Zr2O12} (LLZO), \ce{Li_{0.33}La_{0.56}TiO3} (LLTO), \ce{Li_{1.33}Al_{0.33}Ti_{1.67}(PO4)_{3}} (LATP), and \ce{Li_{3.42}Zn_{0.29}GeO4} (LISICON). Because the precise LIPON composition \ce{Li2PO2N} does not contain accessible Li-vacancies~\cite{2013_Senevirathne_Holzwarth_Solid_State_Ionics}, we generated Li-vacancies by slightly changing the composition to \ce{Li_{1.875}PO_{2.125}N_{0.875}}. In addition, we investigated the Na-SSE material \ce{Na3Zr2Si2PO12} (NASICON), for which we computed also the phase stability window using our own implementation. Table~\ref{tab_materials_computational_details} summarizes the origin of crystallographic files, the corresponding crystal structures, and the supercell compositions used in our computations. 

\begin{table}[t]
\centering
\begin{tabular}{l|c|c|c}
Material & Origin of cif-file & Structure & Supercell composition  \\ \hline
LGPS & ICSD-188886 Ref.~\cite{2013_Kuhn_Lotsch_PCCP} & tetragonal  & \ce{Li40Ge4P8S48} \\
LIPON & Ref.~\cite{2013_Senevirathne_Holzwarth_Solid_State_Ionics} & orthorhombic & \ce{Li45P24O51N21} \\
LLZO & ICSD-422259 Ref.~\cite{2011_Awaka_Akimoto_Chem_Lett} & cubic & \ce{Li56La24Zr16O96} \\
LLTO & ICSD-82671 Ref.~\cite{1996_Fourquet_Crosnier-Lopez_J_Solid_State_Chem} & tetragonal & \ce{Li6La10Ti18O54} \\
LATP & ICSD-253240 Ref.~\cite{2016_Kothari_Kanchan_Physica_B} & hexagonal & \ce{Li16Al4Ti20P36O144} \\
LISICON & ICSD-100169 Ref.~\cite{1978_Hong_Mater_Res_Bull} & orthorhombic & \ce{Li82Zn7Ge24O96} \\
NASICON & ICSD-473 Ref.~\cite{1976_Hong_Mater_Res_Bull} & monoclinic & \ce{Na24Zr16Si16P8O96} \\
\end{tabular}
\caption{Input crystallographic files, structures, and supercells for the investigated SSE materials.}
\label{tab_materials_computational_details}
\end{table}

To validate our implementation of the phase stability method, we also computed the phase stability windows of LGPS and LLZO and compared the results with the published ones~\cite{2015_Zhu_Mo_ACS_Appl_Mater_Interfaces}. For LGPS, we obtained a phase decomposition reduction potential of $\Phi_{\text{red, LGPS}}^{\text{phase}} = (1.97 \pm 0.15)$\,V\textsub{vs.\,Li} and an oxidation potential of $\Phi_{\text{ox, LGPS}}^{\text{phase}} = (1.87 \pm 0.23)$\,V\textsub{vs.\,Li}. Our values for both potentials agree well with the respective values of 1.71 and 2.14\,V\textsub{vs.\,Li} previously reported~\cite{2015_Zhu_Mo_ACS_Appl_Mater_Interfaces}. Within error margins, both potential limits are equal meaning that the phase stability window of LGPS is essentially zero. Errors on the potentials were propagated from an estimated general error of $1.0$\,eV on the total DFT energy of the supercells. Decomposition reactions with errors $>1.0$\,V on the corresponding equilibrium potentials were excluded. Such large errors are encountered for very small $x$ in equation~\eqref{eq_equilibrium_potential}, i.e. for decomposition reactions with very little Li-exchange that are close to the `non-electrochemical' limit. 

For LLZO, we obtained a phase decomposition reduction potential of $\Phi_{\text{red, LLZO}}^{\text{phase}} = (0.02 \pm 0.03)$\,V\textsub{vs.\,Li} and an oxidation potential of $\Phi_{\text{ox, LLZO}}^{\text{phase}} = (2.16 \pm 0.37)$\,V\textsub{vs.\,Li}. Our value for the reduction potential agrees very well with the value of 0.05\,V\textsub{vs.\,Li} previously reported~\cite{2015_Zhu_Mo_ACS_Appl_Mater_Interfaces}. The previously reported value of 2.91\,V\textsub{vs.\,Li} for the oxidation potential is slightly larger than our value, but, given the larger error margin, we still consider it as reasonable agreement. 

\begin{table}[htb]
\centering
\begin{tabular}{l|c|c|c|c|c}
Material & $\Phi_{\text{red}}^{\text{stoi}}$\,/\,V\textsub{vs.\,A} & $\Phi_{\text{ox}}^{\text{stoi}}$\,/\,V\textsub{vs.\,A} & $\Delta\Phi^{\text{stoi}}$\,/\,V & $\Delta E_{\text{HL}}^{\text{N}}$\,/\,eV \\ \hline
LGPS & $\hphantom{+}1.08$ & $2.85$ & $1.77$ & $2.21$ \\
LIPON & $\hphantom{+}0.00$ & $4.11$ & $4.11$ & $5.13$ \\
LLZO & $-0.83$ & $3.52$ & $4.35$ & $4.34$ \\
LLTO & $\hphantom{+}1.65$ & $4.46$ & $2.81$ & $2.56$ \\
LATP & $\hphantom{+}0.66$ & $3.13$ & $2.47$ & $2.48$ \\
LISICON & $-0.32$ & $3.54$ & $3.86$ & $3.63$ \\
NASICON & $-0.17$ & $3.41$ & $3.58$ & $4.34$ \\
\end{tabular}
\caption{Results from DFT computations with PBE functional and SSSP Efficiency pseudopotentials: Lower and upper limits $\Phi_{\text{red / ox}}^{\text{stoi}}$ and corresponding width $\Delta\Phi^{\text{stoi}}$ of the stoichiometry stability windows for the investigated SSE materials. Potential limits are given vs. the reference potential of \ce{A}-metal oxidation (\ce{A}=\ce{Li} for Li-SSE materials, \ce{A}=\ce{Na} for NASICON). Also given are the electronic HOMO--LUMO gaps $\Delta E_{\text{HL}}^{\text{N}}$ of the systems with stable \ce{A}-stoichiometry N.}
\label{tab_stability_results}
\end{table}

\begin{figure}
\begin{center}
\includegraphics[height=7cm]{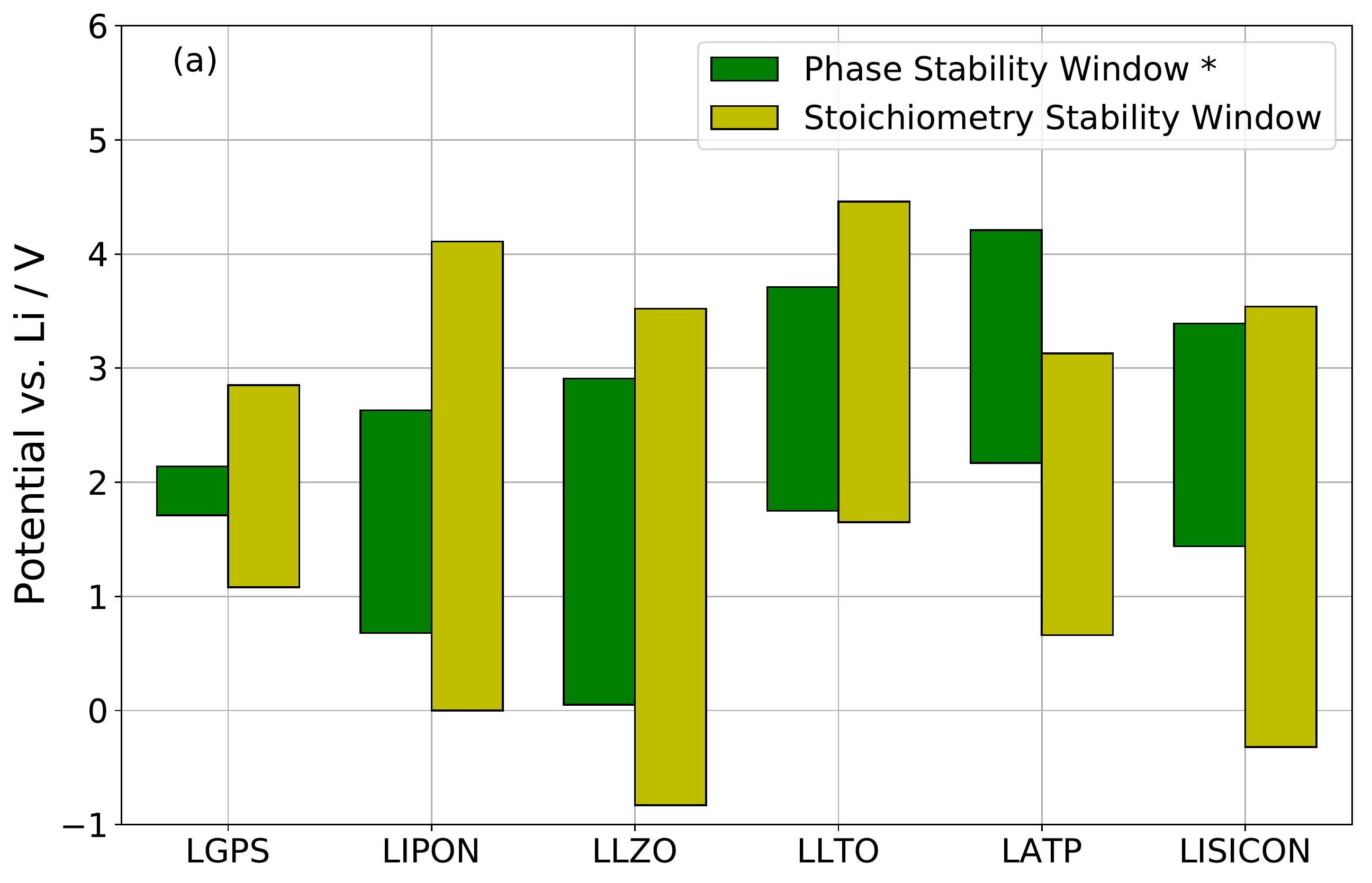}
\includegraphics[height=7cm]{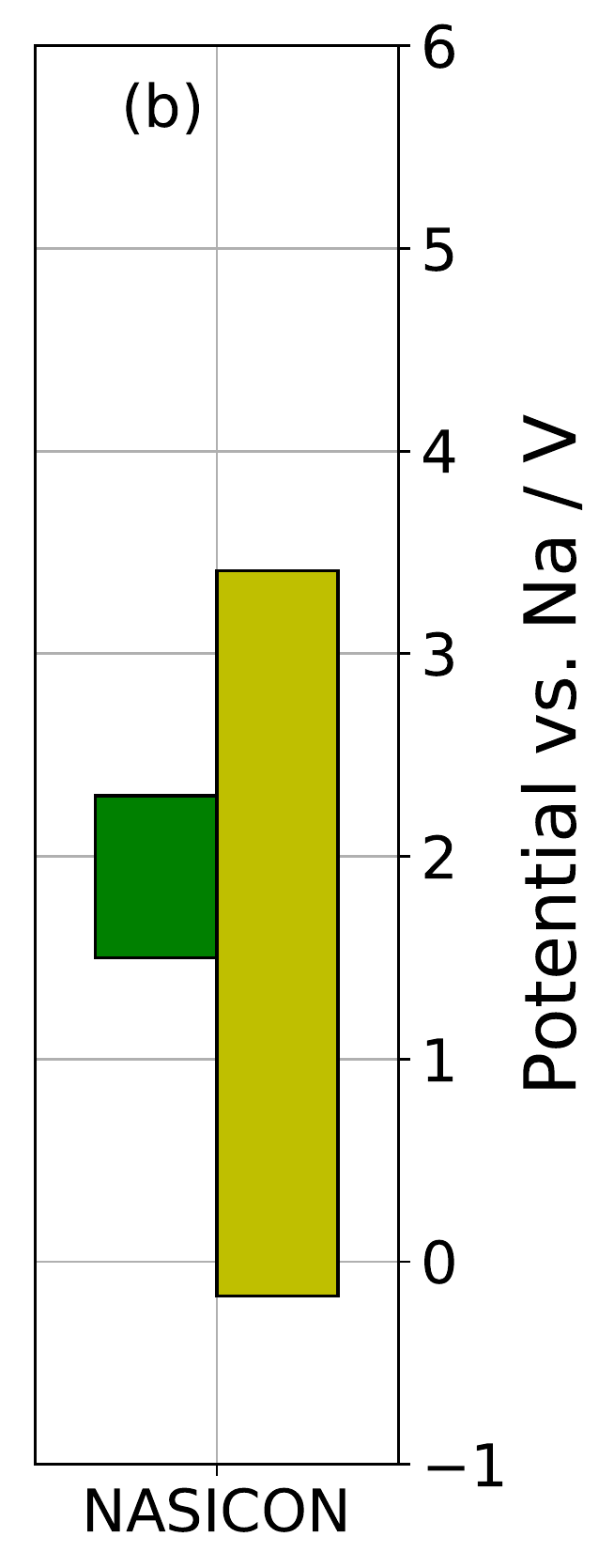} 
\end{center}
\caption{Stoichiometry stability windows and phase stability windows for various Li-SSE materials (a) and NASICON (b). *For Li-SSE materials (a), published phase stability windows from Ref~\cite{2015_Zhu_Mo_ACS_Appl_Mater_Interfaces} are plotted. For NASICON (b) the phase stability window was computed with our own implementation.}
\label{fig_stoichio_phase_stability_results}
\end{figure}

Table~\ref{tab_stability_results} summarizes the computed stoichiometry stability windows for the investigated SSE materials. Our values of the Li-insertion and extraction potentials of LGPS agree well with previously published values~\cite{2016_Bhattacharya_Wolverton_Carbon_Sci_Tech} of 0.78 and 2.97\,V\textsub{vs.\,Li}, respectively. Figure~\ref{fig_stoichio_phase_stability_results}a presents a comparison for the Li-SSE materials with the corresponding phase stability windows published in Ref.~\cite{2015_Zhu_Mo_ACS_Appl_Mater_Interfaces}, which were also computed from DFT energies with PBE xc-functional. The phase stability window of NASICON shown in Figure~\ref{fig_stoichio_phase_stability_results}b was computed using our own implementation of the method. We obtained a phase decomposition reduction potential of $\Phi_{\text{red, NASICON}}^{\text{phase}} = (1.50 \pm 0.23)$\,V\textsub{vs.\,Na} (with decomposition products \ce{ZrO2}, \ce{Na4P2O7}, \ce{P}, and \ce{Na2ZrSi2O7}), and an oxidation potential of $\Phi_{\text{ox, NASICON}}^{\text{phase}} = (2.30 \pm 0.47)$\,V\textsub{vs.\,Na} (with products \ce{ZrO2}, \ce{Na4P2O7}, \ce{Na2ZrSi6O15}, and \ce{O2}), where we included an entropic contribution $-TS = -0.638$\,eV to the free energy of gaseous \ce{O2} at 300\,K and 1\,bar computed from NIST reference data~\cite{NIST_Chem_WebBook}.

In most cases, we find that the phase stability window is more limiting than the stoichiometry stability window, thus supporting the common assumption~\cite{2017_Tian_Ceder_Energy_Environ_Sci}. However, we find an exception for the upper stability limit of LATP, where our value for the stoichiometry oxidation potential is less than the reported phase decomposition oxidation potential~\cite{2015_Zhu_Mo_ACS_Appl_Mater_Interfaces}. Also, in many cases the difference between stoichiometry stability limit and phase stability limit is rather small, especially when taking into account realistic error margins of few hundred meV. This is the case for the lower potential limits of LGPS, LIPON, LLZO, and LLTO and for the upper potential limits of LGPS, LLZO, LLTO, and LISICON. We thus find a correlation between stoichiometry stability window and phase stability window.

LGPS was experimentally found stable at least from 0.0 to 5.0\,V\textsub{vs.\,Li}~\cite{2011_Kamaya_Mitsui_Nature_Mater}, which is significantly wider than both its phase stability window and its stoichiometry stability window. This wide experimental stability window of native LGPS was confirmed in another study~\cite{2015_Han_Wang_Adv_Mater}. After mixing LGPS with carbon, the same authors observed redox processes around 0.0--0.5\,V\textsub{vs.\,Li} and around 1.6--2.7\,V\textsub{vs.\,Li}, respectively. However, we are careful with attributing these potentials either to phase or to stoichiometry stability limits of LGPS, because it appears unclear to us whether the observed redox processes result from LGPS decomposition alone, from carbon, or from an interaction between both. 

Experimental studies of LIPON found an electrochemical stability window from 0.0 to about 5.0--5.5\,V\textsub{vs.\,Li}~\cite{1997_Yu_Hart_J_Electrochem_Soc, 2004_West_J_Power_Sources}, which agrees quite well with our computed stoichiometry stability window. Other authors, however, report decomposition of amorphous LIPON in contact with metallic Li observed by X-ray photoemission spectroscopy (XPS)~\cite{2015_Schwoebel_Solid_State_Ionics}, which is compatible with the positive lower limit of the phase stability window~\cite{2015_Zhu_Mo_ACS_Appl_Mater_Interfaces}. 

Stability of LLZO versus metallic Li was experimentally demonstrated~\cite{2007_Murugan_Weppner_Angewandte,2010_Kotobuki_JECS} in agreement with the computed lower limits $\lessapprox 0.0$\,V\textsub{vs.\,Li} of both phase and stoichiometry stability windows. In one study, LLZO was observed to be stable against oxidation up to very high potentials~\cite{2011_Ohta_J_Power_Sources}. Other authors report an onset of LLZO oxidation at around 4.0\,V\textsub{vs.\,Li}~\cite{2016_Han_Adv_Energy_Mater}, which is very close to the upper limit of our computed stoichiometry stability window, especially from hybrid functional DFT results, cf. Table~\ref{tab_hybrid_results}. However, also in this study carbon was mixed with the LLZO so that other oxidation reactions could also be responsible for the experimental result. 

LLTO is a very interesting case, because the lower limits of both the phase stability window~\cite{2015_Zhu_Mo_ACS_Appl_Mater_Interfaces} and the stoichiometry stability window are almost identical at $\approx 1.7$\,V\textsub{vs.\,Li}. This perfectly agrees with experimental observations: A careful experimental study demonstrated LLTO reduction below 1.7\,V\textsub{vs.\,Li}~\cite{1997_Klingler_Weppner_Ionics}. Because the amount of transferred Li agreed well with the number of available Li vacancies inside the LLTO lattice, insertion of Li into the LLTO host lattice was concluded. At lower potentials, more Li was consumed, which was attributed to the formation of secondary decomposition phases. Thus, this study demonstrated that the stoichiometry stability window defined the lower stability limit of LLTO. A similar LLTO reduction potential was also experimentally determined by other authors~\cite{2001_Chen_Solid_State_Ionics}. 

Another interesting case is LATP, where our computed upper limit of the stoichiometry stability window $\Phi_{\text{ox}}^{\text{stoi}}$ is significantly less than the previously reported phase stability limit $\Phi_{\text{ox}}^{\text{phase}}$~\cite{2015_Zhu_Mo_ACS_Appl_Mater_Interfaces}. This result even holds for our hybrid functional DFT results, cf. Table~\ref{tab_hybrid_results}. Unfortunately, to the best of our knowledge, no experimental study on electrochemical LATP oxidation exists to date that could confirm whether the stoichiometry stability window is indeed defining the upper LATP stability limit.

For LISICON, we find almost identical upper limits of our computed stoichiometry stability window and the previously published phase stability window~\cite{2015_Zhu_Mo_ACS_Appl_Mater_Interfaces}. However, a negative value of the lower limit of the stoichiometry stability window indicates stability against metallic Li, whereas the phase stability window predicts instability. Experimentally, a strong reaction between LISICON and Li metal was observed~\cite{1978_Alpen_Electrochimica_Acta, 2009_Knauth_Solid_State_Ionics}, clearly demonstrating that the lower phase stability limit is critical. 

Turning to the results for NASICON, the negative value of $\Phi_{\text{red}}^{\text{stoi}}$ indicates stoichiometric stability of NASICON against a Na metal electrode, whereas our computed phase decomposition reduction potential $\Phi_{\text{red, NASICON}}^{\text{phase}} = (1.50 \pm 0.23)$\,V\textsub{vs.\,Na} predicts NASICON to be unstable against Na metal in agreement with experimental findings~\cite{1988_Warhus_Rabenau_J_Solid_State_Chem}. However, other authors reported that no reaction was observed between NASICON and metallic Na~\cite{2013_Noguchi_Yamaki_Electrochim_Acta}. Recently, an experimental study reported a very wide electrochemical stability window for Ca-doped NASICON~\cite{2019_Lu_Adv_Energy_Mater}.

The wider experimental stability windows compared to computed phase stability windows observed for many SSE materials were convincingly explained with passivation due to interphase layers of decomposition products between electrode and SSE~\cite{2015_Zhu_Mo_ACS_Appl_Mater_Interfaces}. A complementary possible explanation is a kinetic sluggishness of phase decomposition reactions~\cite{2015_Zhu_Mo_ACS_Appl_Mater_Interfaces}. It was proposed that the \ce{A}-insertion/extraction potentials, i.e. the stoichiometry stability window, provide hard limits for the SSE stability, because the transfer of single \ce{A}-atoms between electrode and SSE is likely to be fast~\cite{2017_Tian_Ceder_Energy_Environ_Sci}. Furthermore, the insertion or extraction of single \ce{A}-atoms into/from the SSE structure might represent the first step in the mechanism of SSE phase decomposition reactions. Therefore, even when strict thermodynamic stability is defined by the phase stability window, the stoichiometry stability window could represent a ``kinetic'' stability window up to which SSE phase decomposition reactions are kinetically hindered. This interpretation could explain why in many cases better agreement is found between experimental stability windows and the stoichiometry stability window rather than the phase stability window.

\begin{figure}
\begin{center}
\includegraphics[height=7cm]{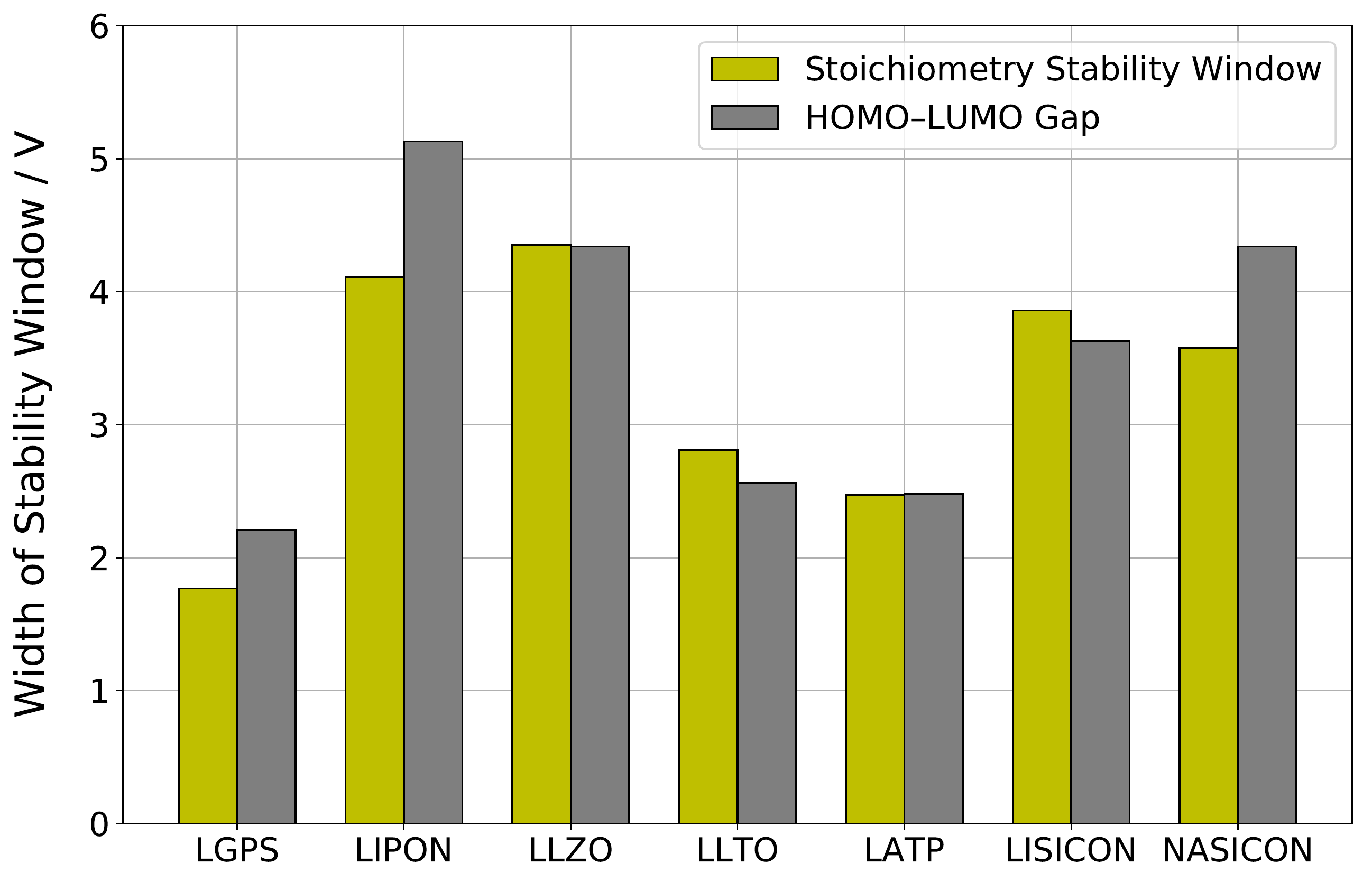} 
\end{center}
\caption{Width $\Delta\Phi^{\text{stoi}}$ of the stoichiometry stability windows in comparison with the HOMO--LUMO gap $\Delta E_{\text{HL}}^{\text{N}}$ for the investigated SSE materials.}
\label{fig_stoichio_bandgap_stability_results}
\end{figure}

\paragraph{Comparison of stoichiometry stability window and HOMO--LUMO gap.}
The electronic HOMO--LUMO gaps $\Delta E_{\text{HL}}^{\text{N}}$ of the systems with stable \ce{A}-stoichiometry N are compiled in Table~\ref{tab_stability_results}. Figure~\ref{fig_stoichio_bandgap_stability_results} presents a comparison of the widths $\Delta\Phi^{\text{stoi}}$ of the stoichiometry stability window with the HOMO--LUMO gaps $\Delta E_{\text{HL}}^{\text{N}}$ for the investigated SSE materials. As expected from equation~\eqref{eq_width_stoi_window_IP_EA} and the discussion thereafter, a very good agreement is observed for LGPS, LLZO, LLTO, LATP, and LISICON. Only for LIPON and NASICON, $\Delta E_{\text{HL}}^{\text{N}}$ is significantly larger than $\Delta\Phi^{\text{stoi}}$. The latter discrepancies can result from electronic states of the conduction band that are shifted down into the band gap upon insertion of the additional \ce{A^+} ion in the N+1 stoichiometry system.

The HOMO--LUMO method~\cite{2010_Goodenough_Kim_Chem_Mater, 2011_Ong_Ceder_Chem_Mater, 2012_Mo_Ceder_Chem_Mater} considers only electron transfer between electrode and SSE. Consequently, the electrochemical stability window is assessed from the ionization potential and electron affinity of the SSE. Transfer of \ce{A^+} ions is neglected and the electrode is considered to be chemically inert. Therefore, the HOMO--LUMO gap is generally considered to provide only an upper bound for the true stability potential window of an SSE material~\cite{2012_Mo_Ceder_Chem_Mater, 2017_Lu_Ciucci_Chem_Mater}. However, according to relation~\eqref{eq_width_stoi_window_IP_EA} and supported by our computational results, the HOMO--LUMO gap is essentially identical to the width of the stoichiometry stability window. Therefore, both methods implicitly consider the same physical process, i.e. an infinitesimal exchange of \ce{A} between electrode and SSE. Consequently, as discussed above, exceptional cases can exist where the limiting stability window of an SSE material is defined by its stoichiometry stability potentials, i.e. its HOMO--LUMO gap. The advantage of the stoichiometry stability method over the HOMO--LUMO method is that the former directly yields the absolute position of the stability window vs. a reference potential. 

\begin{figure}
\begin{center}
\includegraphics[height=7cm]{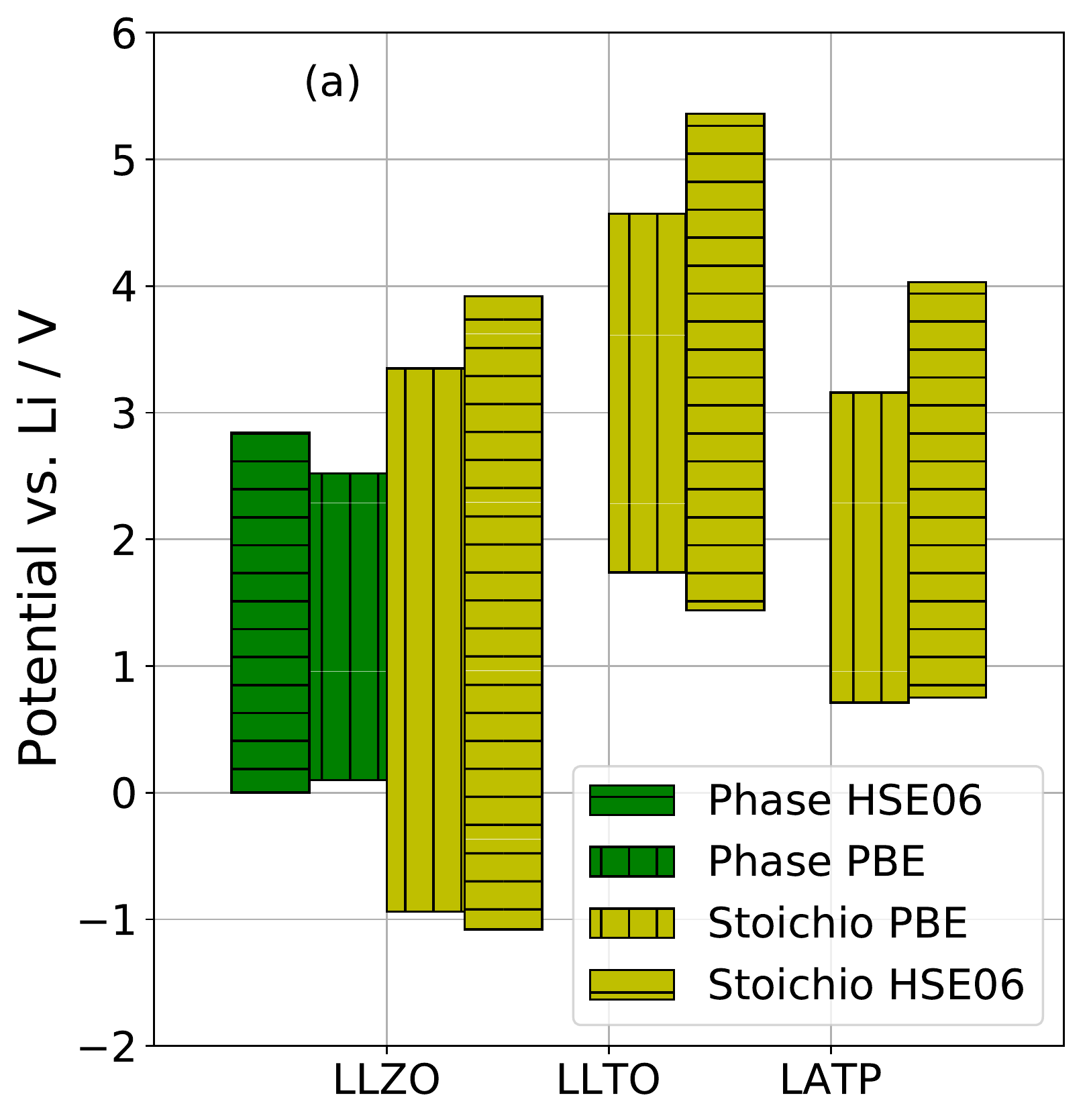}
\hspace{0.2cm} 
\includegraphics[height=7cm]{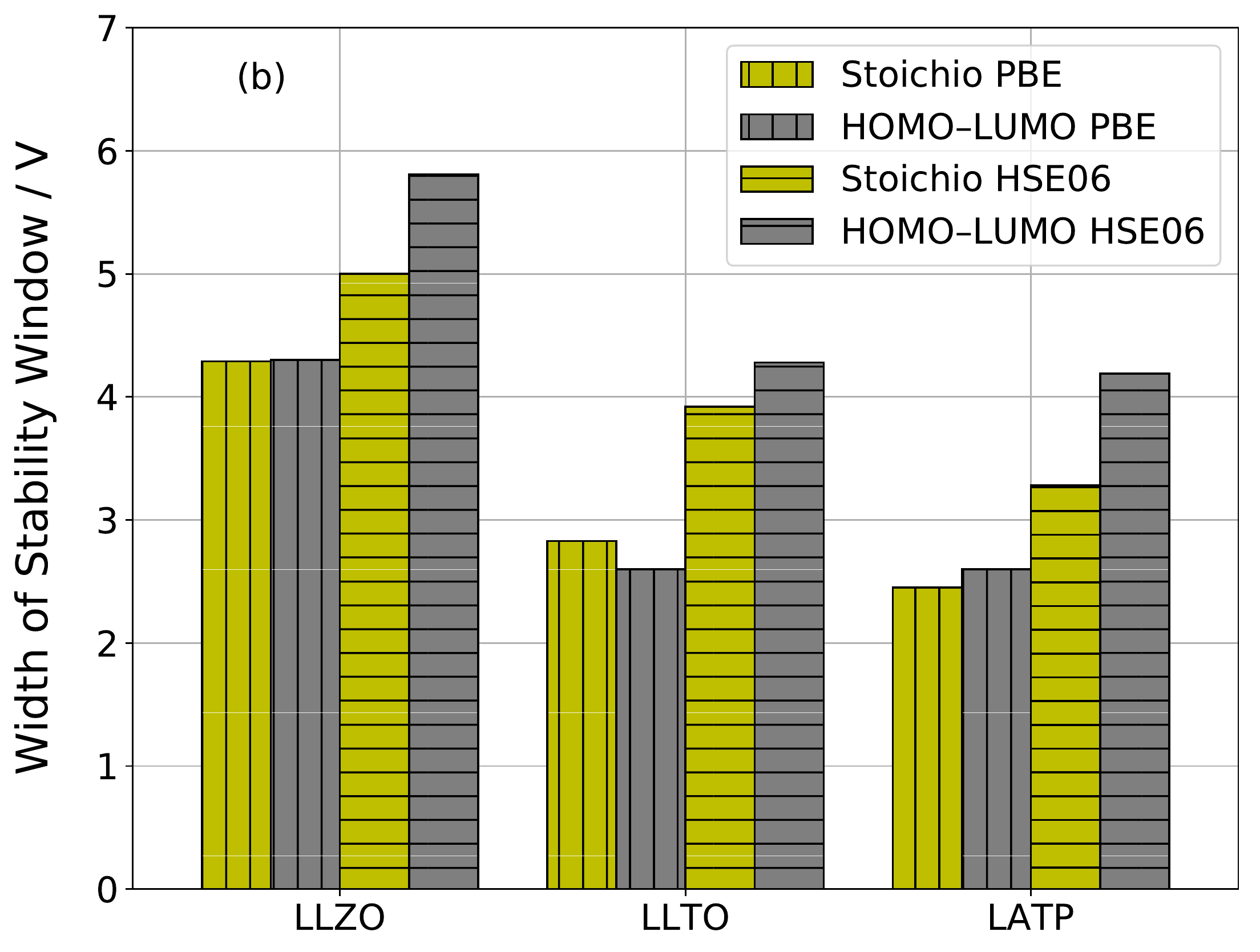} 
\end{center}
\caption{Comparison of stability windows computed with PBE and with HSE06 DFT functionals. (a) Phase stability windows and stoichiometry stability windows. (b) Width $\Delta\Phi^{\text{stoi}}$ of the stoichiometry stability windows and HOMO--LUMO gaps $\Delta E_{\text{HL}}^{\text{N}}$.}
\label{fig_hybrid_results}
\end{figure}

\paragraph{Influence of hybrid functional DFT.}
We calculated the phase stability window of LLZO at the hybrid functional DFT level. Using the HSE06 functional, we recomputed only the reaction energies of the limiting reduction and oxidation decomposition reactions that were obtained from our implementation of the phase stability method with PBE functional. The respective decomposition products are \ce{Li2O}, \ce{La}, \ce{Li6Zr2O7} for the limiting reduction reaction, and \ce{Li2O2}, \ce{La2O3}, \ce{Li6Zr2O7} for the limiting oxidation reaction. We obtained $\Phi_{\text{red, LLZO}}^{\text{phase, HSE06}} = 0.00$\,V\textsub{vs.\,Li} and $\Phi_{\text{ox, LLZO}}^{\text{phase, HSE06}} = 2.84$\,V\textsub{vs.\,Li}. Also, we recomputed the stoichiometry stability windows of LLZO, LLTO, and LATP using the HSE06 hybrid functional for the minimum energy configurations obtained from the stoichiometry stability workflow with PBE functional. Results are summarized in Table~\ref{tab_hybrid_results} and presented in Figure~\ref{fig_hybrid_results}a. Note that we had to use different pseudopotentials (SG15 ONCV) for hybrid DFT computations than the SSSP Efficiency pseudopotentials used in the phase stability and stoichiometry stability workflows. For direct comparability, we also recomputed the critical PBE DFT energies with the SG15 ONCV pseudopotentials for the PBE results shown in Figure~\ref{fig_hybrid_results}a, which, therefore, slightly differ from the results presented in Figure~\ref{fig_stoichio_phase_stability_results}. 

We find that the hybrid functional has only minor influence on the reduction potential limits of the investigated phase stability window and stoichiometry stability windows. In these cases, a reliable prediction of SSE stability against a metallic Li electrode is obtained at PBE level. However, PBE results significantly underestimate the oxidation potential limits of the stoichiometry stability windows compared to HSE06. Also the oxidation potential limit of the LLZO phase stability window is increased using HSE06. Thus, the computation of reliable oxidation potentials appears more demanding than for reduction potentials. This difference might originate from the difference between anionic vs. cationic redox behaviour, because SSE reduction largely corresponds to cationic reduction processes, whereas SSE oxidation can affect both cation and anion valencies.

\begin{table}[htb]
\centering
\begin{tabular}{l|c|c|c|c|c|c|c}
Material & $\Phi_{\text{red}}^{\text{stoi}}$\,/\,V\textsub{vs.\,Li} & $\Phi_{\text{ox}}^{\text{stoi}}$\,/\,V\textsub{vs.\,Li} & $\Delta\Phi^{\text{stoi}}$\,/\,V & $\Delta E_{\text{HL}}^{\text{N}}$\,/\,eV \\ \hline
LLZO & $-1.08$ & $3.92$ & $5.00$ & $5.81$ \\
LLTO & $\hphantom{+}1.44$ & $5.36$ & $3.92$ & $4.28$ \\
LATP & $\hphantom{+}0.75$ & $4.03$ & $3.28$ & $4.19$ \\
\end{tabular}
\caption{Results from hybrid functional DFT computations with HSE06 functional and SG15 ONCV pseudopotentials: Lower and upper limits $\Phi_{\text{red / ox}}^{\text{stoi}}$ and corresponding width $\Delta\Phi^{\text{stoi}}$ of the stoichiometry stability windows, and the electronic HOMO--LUMO gaps $\Delta E_{\text{HL}}^{\text{N}}$ of the systems with stable \ce{A}-stoichiometry N.}
\label{tab_hybrid_results}
\end{table}

Figure~\ref{fig_hybrid_results}b presents a comparison of the widths of stoichiometry stability windows and HOMO--LUMO gaps obtained with PBE and HSE06 functionals for LLZO, LLTO, and LATP. In all cases, both the width of the stoichiometry stability window and the HOMO--LUMO gap increase going from PBE to HSE06, which is expected because GGA functionals are known to underestimate band gaps whereas hybrid functionals yield better estimates~\cite{2017_Perdew_PNAS}. Surprisingly, however, the agreement between the widths of stoichiometry stability windows and the HOMO--LUMO gaps is significantly reduced at HSE06 level. 

\section{Conclusion}

We analysed the relation between three different methods to compute the electrochemical stability window of solid-state electrolytes, namely the HOMO--LUMO method, the phase stability method, and the stoichiometry stability method. Interestingly, the latter method represents a link between the former two. The width of the stoichiometry stability window is equal to the fundamental gap of the SSE. Whereas the phase stability method takes into account the equilibrium potentials of SSE phase decomposition reactions, the stoichiometry stability method considers the insertion and extraction of single \ce{A}-atoms (\ce{A} = Li, Na) into/from the intact SSE structure. Because these sets of instability reactions are disjoint, the corresponding phase and stoichiometry stability windows are complementary. We further provided computational implementations of the methods and we compared the results for the relevant Li- and Na-SSE materials LGPS, LIPON, LLZO, LLTO, LATP, LISICON, and NASICON. In most cases, the phase stability window is stricter than the stoichiometry stability window, but we also found exceptions to that rule. Comparison with published experimental stability windows revealed an ambiguous picture: Whereas for some SSE materials the experimental observations agree with computational phase stability windows, for other SSE materials experimental stability windows are wider and in better agreement with the computational stoichiometry stability windows. 

\section*{Conflicts of interest}
There are no conflicts to declare.

\begin{acknowledgement}
This research was supported by the NCCR MARVEL, funded by the Swiss National Science Foundation. This work was supported by a grant from the Swiss National Supercomputing Centre (CSCS) under project IDs mr0, mr18 and mr28. We thank Leonid Kahle from EPFL, Lausanne, Switzerland, for support in the computational implementation. We also thank Dr. Nicola Colonna from PSI, Villigen, Switzerland, for helpful discussions regarding hybrid functional DFT.
\end{acknowledgement}

\section*{Electronic Supplementary Information (ESI) available}
Derivation of Nernst equation; Simple model for stoichiometry stability window; Extended model for stoichiometry stability window.

\bibliography{stability_potential_window_bibliography}

\end{document}